\documentclass[12pt]{iopart}

\usepackage{iopams}     
\usepackage{graphicx}
\usepackage{epsfig}
\usepackage{color}
\usepackage{url}
\usepackage{times}
\usepackage{bm}         
\usepackage{times}

\newcommand{\beq}{\begin{equation}}
\newcommand{\eeq}{\end{equation}}
\newcommand{\beqn}{\begin{eqnarray}}
\newcommand{\eeqn}{\end{eqnarray}}

\usepackage{color}

\usepackage{graphicx}
\usepackage{dcolumn}
\usepackage{bm}
\usepackage{epsf}

\begin{document}

\title{Equation of state effects in black hole-neutron star mergers}

\author{Matthew D. Duez$^1$, Francois Foucart$^1$, Lawrence E. Kidder$^1$,
Christian D. Ott$^2$, and Saul A. Teukolsky$^{1,2}$}

\address{$^1$ Center for Radiophysics and Space
    Research, Cornell University, Ithaca, New York, 14853 \\
$^2$ Theoretical Astrophysics 350-17,
    California Institute of Technology, Pasadena, CA 91125}

\begin{abstract}
The merger dynamics of a black hole-neutron star (BHNS) binary is
influenced by the neutron star equation of state (EoS) through the
latter's effect on the neutron star's radius and on the character of
the mass transfer onto the black hole.  We study these effects by
simulating a number of BHNS binaries in full general relativity using
a mixed pseudospectral/finite difference code.  We consider several
models of the neutron star matter EoS, including $\Gamma=2$ and
$\Gamma=2.75$ polytropes and the nuclear-theory based Shen EoS.  For
models using the Shen EoS, we consider two limits for
the evolution of the composition: source-free advection and instantaneous
$\beta$-equilibrium.  To focus on EoS effects, we fix the
mass ratio to 3:1 and the initial aligned black hole spin to $a/m=0.5$
for all models.  We confirm earlier studies which found that more
compact stars create a stronger gravitational wave signal but a
smaller postmerger accretion disk.  We also vary the EoS while holding
the compaction fixed.  All mergers are qualitatively similar, but we
find signatures of the EoS in the waveform and in the tail and disk
structures.
\end{abstract}



\section{Introduction}
\label{intro}
Black hole-neutron star (BHNS) inspirals and mergers are excellent
gravitational wave sources.  Also, a BHNS merger may leave a hot,
massive accretion disk around the black hole (BH), a promising setup
for producing a short-duration gamma-ray burst
(GRB).  However, reliable predictions regarding the waveform and
post-merger state of any specific BHNS merger can only be obtained
from fully relativistic numerical simulations.

The merger is strongly affected by the NS equation of
state (EoS), i.e., the dependence of the fluid pressure on density,
temperature, and composition.  The EoS determines the radius
$R_{\rm NS}$ of a NS of given mass $M_{\rm NS}$, and thus it also
fixes the star's compaction $\mathcal{C}=M_{\rm NS}/R_{\rm NS}$. 
A larger star will be tidally disrupted by the BH's
gravity at a larger distance.  Also, because the EoS affects
$dR_{\rm NS}/dM_{\rm NS}$, it can influence the character of the mass transfer
from the star to the BH, with mass transfer tending to be more stable
for stiffer EoS.  Differences in the stability of mass transfer could
conceivably result in qualitative differences in the merger dynamics. 
The NS EoS at low densities is known, but at high densities it is
not well constrained by theory, experiment, or observation.  It must, therefore, be treated as
another parameter in BHNS simulations.  If the EoS significantly affects
the waveform, then comparison of numerically generated waves with observations may
someday provide information about the EoS.

Attempts have been made to estimate the effects of NS EoS in the context
of Newtonian BHNS simulations.  Newtonian simulations have considered
polytropic~\cite{Lee:1998qk,Lee:1999kcb,Lee:2000uz,Lee:2001ae},
Lattimer-Swesty~\cite{Lattimer:1991nc,Janka:1999qu} and
Shen~\cite{Shen:1998gq,1998PThPh.100.1013S,Rosswog:2004zx} EoS. 
These simulations showed
large qualitative differences for different EoS assumptions.  For
Lattimer-Swesty nuclear matter, the NS disrupts in one mass transfer event,
and a large post-merger disk is created.  For Shen nuclear matter, a NS core
can survive multiple mass transfer events, and the postmerger disk is much
smaller.  There are indications that the differences are not quite so
dramatic when general relativity (GR) is accounted for.  The use of
GR-mimicking potentials~\cite{1980A&A....88...23P,1996ApJ...461..565A} 
tends to eliminate episodic mass transfer~\cite{2005ApJ...634.1202R,
Ruffert:2009um}.  Simulations of large mass-ratio cases using the conformally
flat approximation of GR also found mass transfer to be less stable
(and, therefore, surviving cores to be less likely) than in Newtonian
evolutions~\cite{Faber:2005yg}.  These conformally-flat-GR studies only considered
fairly soft EoS, however.

Published BHNS simulations in full GR have thus-far restricted
themselves to modeling NS as $\Gamma=2$
polytropes~\cite{Shibata:2006ks,
  Shibata:2006bs,Etienne:2007jg,Shibata:2007zm,Duez:2008rb,Etienne:2008re,
  Shibata:2009cn}.  These have investigated the effects of varying one
important EoS-related variable, namely $R_{\rm
  NS}$~\cite{Etienne:2007jg,Shibata:2007zm,Shibata:2009cn}.  Varying
$R_{\rm NS}$ between 12 and 15km for a fixed $M_{\rm BH}$ and $M_{\rm
  NS}$, they find, unsurprisingly, that larger stars disrupt farther
from the BH, leading to larger disks but weaker merger waveforms.
These simulations could not address the importance of the stiffness
of the EoS in affecting the mass transfer.

Another important question is whether BHNS mergers eject significant amounts
of NS matter (perhaps including r-process elements~\cite{1974ApJ...192L.145L,
2008ApJ...679L.117S}) into the interstellar medium. 
Newtonian~\cite{Janka:1999qu},
pseudo-Newtonian~\cite{2005ApJ...634.1202R, Ruffert:2009um}, and
conformal-GR~\cite{Faber:2005yg} simulations predict large ejecta
masses, but this has not yet been seen in full GR.
However, microphysics in the tidal tail could have important effects.
As the density decreases, nucleons recombine into nuclei, heating the
gas through the released binding energy.  The resulting increase in
thermal pressure can strongly affect the tail~\cite{Rosswog:2004zx},
perhaps helping to unbind material.  Material may also be ejected in
a neutrino-driven or magnetic field-driven disk wind, effects which are also
not captured in current GR simulations.

In this paper, we study the imprint of the NS EoS on the merger
dynamics and the resulting waveform and disk.  Our investigation is
based on simulations of BHNS mergers in full GR carried
out with the Cornell-Caltech code SpEC.
In a previous paper~\cite{Duez:2008rb}, 
we demonstrated this code's ability to evolve BHNS inspirals and mergers. 
Since then, we have made a number of improvements which reduce
the constraint violation during merger by an order of magnitude. 
For this study, we fix the binary mass ratio to 3:1 and the initial BH spin to be
$\left|\vec{S}_{\rm BH}\right|/M_{\rm BH}{}^2=0.5$ (aligned with the
orbital angular momentum), and we vary the assumed EoS. 
We evolve $n=1$ polytropes using a $\Gamma=2$ Gamma-law and $n=4/7$ polytropes
using $\Gamma=2.75$.  We also perform simulations using the tabulated, nuclear-theory
based Shen EoS~\cite{Shen:1998gq,1998PThPh.100.1013S}.  We vary the EoS
and the compaction separately to isolate the effect of each. 

All our mergers are qualitatively similar:  we find no cases of episodic
mass transfer, no instances of measurable unbounded outflow, and no cases
in which the disruption fails to produce a significant disk.  There are,
however, some quantitative EoS signatures in the waveform above 1kHz and in the disk. 
We confirm earlier findings regarding the effects of 
NS compaction.  For a given compaction, stiffer EoS produce larger, longer-lived tidal tails. 
In every case, the post-merger disk has a mass
of 0.05-0.1$M_{\odot}$ and an average temperature of order an MeV.

In Section 2, we describe
important improvements to our code from~\cite{Duez:2008rb}. 
In Section 3, we describe the
models we evolve.  In Section 4, we present the results of our evolutions.  In Section 5,
we summarize our findings and consider
what remains to be done to accurately sample all of the interesting regions of
BHNS parameter space.

\section{Numerical methods}
\label{methods}

\subsection{Code improvements}
\label{improvements}

The main improvements to our code from~\cite{Duez:2008rb} come from a better allocation of grid points
when solving the fluid equations, modified gauge conditions and atmosphere prescriptions.

Our simulations use two grids~\cite{Duez:2008rb}: a pseudospectral grid on which Einstein's equations
are solved, and a finite difference (FD) for the relativistic fluid equations. For efficient evolutions, 
the FD grid should adapt to the configuration of the fluid. One approach
is to use adaptive mesh 
refinement, as in some FD-based GR codes~\cite{Schnetter:2003rb,Anderson:2006ay,Yamamoto:2008js}. 
We choose instead to modify the mapping between the two grids as soon as
a significant flow of matter is approaching the boundary of the FD grid. A translation and
rescaling of the coordinates allows the grid to move, grow, and shrink so as to follow the fluid 
evolution. To increase the resolution in the neighborhood of the black hole, we use a map of the form
\begin{eqnarray}
\label{eq:regrid}
r' & = & r {\;\;\;(r<R_0)}  \nonumber \\
r' & = & ar^3+br^2+cr+d {\;\;\; (R_0<r<R_1)}\\
r' & = & \alpha r {\;\;\;(r>R_1)}, \nonumber
\end{eqnarray}
 where $r$ is the coordinate distance to the center of the hole, $R_0$ and $R_1$ are predetermined length scales, the parameters $(a,b,c,d)$ are chosen so that
the map is $C^1$, and $\alpha$ varies with the size of the FD grid. (Once the grid becomes large, we maintain a constant resolution in the region 
$r<R_0$ and vary $\alpha$ to fix the location of the outer boundaries.)

The gauge in the generalized harmonic formulation is set by specifying the functions
$H_a=g^{bc}\Gamma_{abc}$.  During the inspiral, we fix $H_a$ in the
moving frame, as in our earlier paper~\cite{Duez:2008rb}.  During the merger, we continue
to hold $H_a$ fixed in the moving frame in the region near the black hole, and we
exponentially damp $H_a$ to zero in the region far from the hole.  We find that fixing $H_a$ near
the excision zone significantly reduces constraint violations.

As Faber~{\it et al.}~\cite{Faber:2007dv} have pointed out, the inversion from
conservative to primitive hydrodynamic variables is only possible if
$S \equiv g^{ij}S_iS_j < S_{\rm max} \equiv \tau (\tau + 2D)$.  
After each evolution step, we impose the condition
$S\leq S_{\rm max}^{\rm code} = f S_{\rm max}$.  We find $0.99<f<1$ is necessary
to avoid causing large effects on the evolution of the tidal tail. 
Once the values are ``fixed'' in this way, primitive variables can
be reconstructed, but they may still be unreasonable in very low
density ``atmosphere'' regions.
Therefore, we next apply limits on the conformal 3-velocity
$u_i$ and the temperature $T$.  We emphasize that these
limits are only applied to low density regions, several orders of
magnitude sparser than the star, the disk, or the tidal tail.  We have checked
that our evolutions are insensitive to variations in these atmosphere ceilings.

With our standard resolution, the normalized constraint violations
peak at $\approx1$\% during mergers.  Convergence tests on the $\Gamma=2.75$ case
indicate errors in our reported disk masses of $<10$\%.

\subsection{Use of tabulated $\rho/T/Y_e$-dependent EoS}
For composition-dependent EoS, there is a new independent variable to
be evolved:  the electron fraction $Y_e$.  Its evolution equation
in conservative form is
\begin{equation}
  \partial_t(DY_e) + \partial_i(DY_ev^i) = S_{\nu}\ ,
\end{equation}
where $S_{\nu}$ is the source term set by weak interactions and
neutrino radiation, effects not modeled in our code.
Here, we consider two limiting cases.  First, we
assume the weak interaction timescales are much longer than the merger
timescale.  Then we may set $S_{\nu}=0$ and evolve a continuity equation
for $DY_e$.  As another limit, we assume that weak interactions act
sufficiently quickly to instantaneously enforce $\beta$-equilibrium.  Thus,
for a given density $\rho$ and $T$, $Y_e$ is set to the value that makes
$\mu_n = \mu_p + \mu_e$, where $\mu_x$ is the chemical potential of particle $x$. 
(We assume $\mu_{\nu}$ is negligible.)  This effectively removes $Y_e$ as a
dynamical variable.  In neither case do we account
for energy loss by neutrino emission.  Given that
the cooling timescale of the disk will probably not be less than about
0.1 second (see, e.g.~\cite{Lee:2005se}), and our simulations last $\sim$10ms,
ignoring neutrino cooling is reasonable. 

We use a tabulated EoS with baryon component taken from
Shen~{\it et al}~\cite{Shen:1998gq,1998PThPh.100.1013S} and with
lepton and photon contributions added.  (See~\cite{2009arXiv0912.2393O}
for details.)  
For low $T$ ($\le 10$MeV), the NS EoS is fairly soft ($\Gamma\approx 4/3$)
at low $\rho$ ($\lesssim 10^{12}{\rm g\ cm^{-3}}$) and stiffer at high
$\rho$ ($\Gamma\approx 2.75$ in the Shen model).
The Shen EoS predicts a TOV maximum gravitational mass of 2.2$M_{\odot}$.

\section{Cases}
\label{cases}
In Table~\ref{table:init}, we present the initial data for the cases we
evolve.  For this study, we do not consider the effects of the binary
mass ratio $q=M_{\rm BH}/M_{\rm NS}$ and set $q=3$ throughout. For
each case, we set the NS baryonic mass to 1.55$M_{\odot}$.  We wish
to study cases that lead to massive disks, and so we include in each case
an initial BH spin of $s\equiv \left|\vec{S}_{\rm BH}\right|/M_{\rm BH}^2=0.5$
orthogonal to the orbital plane. 
Finally, we neglect the NS spin altogether and focus solely on what is thought
to be a good approximation for the most likely scenarios:  irrotational
stars~\cite{1992ApJ...400..175B,1992ApJ...398..234K}.

To study the effect of compaction, we ran two cases with $\Gamma=2.75$, one
with $\mathcal{C}=M_{\rm NS}/R_{\rm NS}=0.146$ and one with $\mathcal{C}=0.173$. 
These runs are labeled ``$\Gamma$2.75c.15'' and ``$\Gamma$2.75c.17'' in the
tables and figures below.  For our chosen rest mass, the NS gravitational mass
is about $M_{NS}=1.4M_{\odot}$, so the two compactions correspond to radii of
$14.4$km and $12.1$km.

To separate the effects of the EoS from those of the compaction, we use three
equations of state with the same $R_{\rm NS}$ ($\mathcal{C}=0.15$).  This comparison
has not been attempted in previously published full-GR numerical studies. 
We use two $\Gamma$-law EoS:  $\Gamma=2$ (run ``$\Gamma2$'') and $\Gamma=2.75$. 
We also use the tabulated Shen EoS.  As mentioned above, we evolve the initial data
with Shen EoS in two ways:  assuming instantaneous $\beta$-equilibrium (run ``Shen-$\beta$'')
and assuming $S_{\nu}=0$ (run ``Shen-Adv'').

All the initial configurations are generated using our multidomain spectral
elliptic solver~\cite{Pfeiffer:2003a} to solve for quasi-equilibrium configurations of
BHNS binaries in the extended conformal thin sandwich formalism~\cite{Foucart:2008a}.
We do not assume conformal flatness, but instead choose a conformal metric
approaching Kerr in the neighborhood of the BH~\cite{Lovelace:2008a,Foucart:2008a}. 
For the polytropic runs, the initial temperature $T_{\rm init}$ is zero.  For the Shen runs,
$T_{\rm init}=0.1$MeV, and the initial $Y_e$ is set by assuming $\beta$-equilibrium.
To ensure that the comparison is not affected by the initial eccentricity of the binary,
we also apply the eccentricity removal technique devised by Pfeiffer et al.~\cite{Pfeiffer:2007a}
until $e\sim0.01$. (Without eccentricity removal, we would have $e=0.01-0.06$.)

\Table{\label{table:init}
Initial data for all runs. $d_{\rm init}$ is the initial coordinate separation and
$\rho^{\rm c}_{14}$ is the central density of the star in units of $10^{14}{\rm g\ cm^{-3}}$
assuming a star of baryonic mass $M_{b}^{\rm NS}=1.55 M_{\odot}$.
$e=\frac{B}{\omega d}$ is the eccentricity and $B$ is obtained by fitting
the evolution of the trajectory to
$\dot{d}=A_0+A_1 t + B \sin(\omega t+\phi)$.  $M$ is the ADM mass of the system. }
\br
Case & EoS & $s_{\rm init}$ & $q$ & $\mathcal{C}_{\rm init}$ & $\Omega_{\rm init}M$ & $d_{\rm init}/M$ &  $\rho^{\rm c}_{14}$ & $e$\\ 
\mr
$\Gamma$2       & $\Gamma=2.00$ & 0.5 & 3 & 0.144 & 0.041 & 7.48 & 7.1 & 0.006\\
$\Gamma$2.75c.15    & $\Gamma=2.75$ & 0.5 & 3 & 0.146 & 0.041 & 7.45 & 4.5 & 0.003\\
$\Gamma$2.75c.17 & $\Gamma=2.75$ & 0.5 & 3 & 0.173 & 0.041 & 7.42 & 7.6 & 0.014\\
Shen-$\beta$       & \rm{Shen} & 0.5 & 3 & 0.147 & 0.041 & 7.59 & 5.0 & 0.004\\
Shen-Adv           & \rm{Shen} & 0.5 & 3 & 0.147 & 0.041 & 7.59 & 5.0 & 0.004\\
\br
\endTable

\section{Results}

\subsection{Qualitative features of the mergers}
\label{feature}

For each configuration, the binary passes through 2 -- 3 orbits of
inspiral before reaching a coordinate separation of $d\approx$40km (5$M$), at
which point matter starts flowing from the star to the BH.  Most of
the core of the star is then rapidly accreted onto the hole [within
$t\sim$3ms (100$M$)] while a large tidal tail forms.  For every EoS, even the
stiffest ($\Gamma=2.75$), the star is disrupted in one extended mass
transfer event.  This confirms earlier indications that episodic mass
transfer does not happen in GR for the realistic range of neutron star
EoS.  All tails but the one formed by the most compact star extend to
distances $d\geq$400km.  All of this material, however, remains bound to
the system.
This result is significant because our Shen-EoS runs do account for
recombination effects that occur for $\rho<10^{14}{\rm g\ cm}^{-3}$. 
It would seem that it is the
inclusion of GR, and not inferior microphysics, that causes GR
simulations to see no ejecta.  (It is, however, possible that some
very low-density material is ejected and that this is suppressed in
our simulations by the atmosphere prescription.)

Material from the tail falling back
towards the hole has enough angular momentum to avoid being immediately accreted: once the core of the
star has fallen into the black hole, the accretion rate drops and a disk forms from the remains of the
tail. This disk is at first neither axisymmetric nor time-independent:  the density peaks strongly at the
junction of the disk and the tidal tail, and the matter keeps expanding away from the hole.
As the accretion rate decreases, a gap opens between the disk and the black hole.   At the end of our simulations, the outer disk is still quite nonaxisymmetric, and the tidal
tail is settling back onto the disk.  The inner disk has a density
$\rho\sim 10^{11}-10^{12} {\rm \:g\:cm^{-3}}$ and temperature $T\sim 0.1-10{\rm \:MeV}$. It is also relatively
thick ($H_{\rm disk}/R_{\rm disk}\sim0.1-0.3$).

\Table{\label{table:final_BH}
The properties of the BH and accretion torus at time $t=t_{\rm merger}+8.3$ms.
$v_{\rm kick}$ is the BH kick velocity in km s${}^{-1}$. 
$M_{\rm disk}$ is the baryonic mass outside of the black hole,
$r_{\rm disk}$ is the radial extent of the disk,  
$\rho^{\rm max}_{\rm disk,12}$ is the maximum density in units of
$10^{12}$g~cm${}^{-3}$,  
$\langle T\rangle_{\rm disk}$ is the density-weighted average temperature in MeV
and $\Psi_4^{2,2}$ is the amplitude of gravitational waves in the (2,2) mode
extracted at $r=75M$.
Note that the disk continues to evolve at late times as
$r_{\rm disk}$ and $\langle T\rangle_{\rm disk}$ slowly increase.}
\br
Case & $M_{\rm BH}/M$ & $s_{\rm final}$ & $v_{\rm kick}$ & $M_{\rm disk}/M_{\rm NS}$ & $r_{\rm disk}/M$ & $\rho^{\rm max}_{\rm disk,12}$ & $\langle T\rangle_{\rm disk}$  & $rM\Psi_4^{2,2}$\\ 
\mr
$\Gamma$2          & 0.96 & 0.69 & 83 & 0.08  & 12 & 1   & 1.3 & 0.012\\
$\Gamma$2.75c.15   & 0.94 & 0.69 & 45 & 0.13  & 12 & 1   & 0.6 & 0.010 \\
$\Gamma$2.75c.17   & 0.96 & 0.70 & 75 & 0.02  & 6  & 0.3 & 1.2 & 0.030 \\
Shen-$\beta$       & 0.97 & 0.79 & 63 & 0.07  & 10 & 1   & 2.7 & 0.015 \\
Shen-Adv           & 0.97 & 0.80 & 61 & 0.07  & 11 & 1   & 2.5 & 0.016 \\
\br
\endTable

\subsection{Effects of composition, compaction, and stiffness}
\label{parameter_study}

\begin{figure}
\begin{center}
\includegraphics[width=9cm]{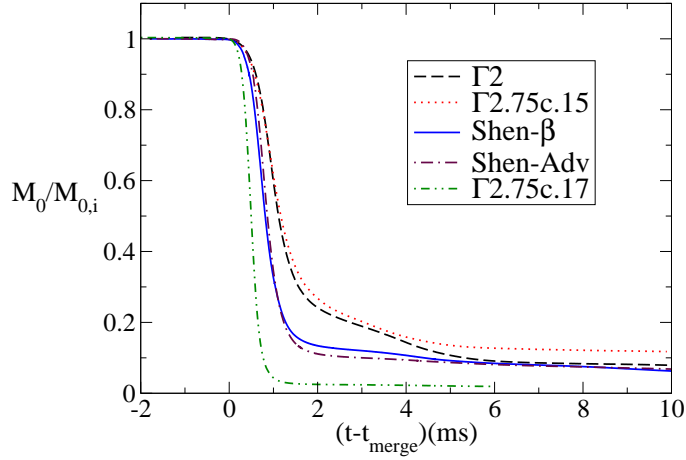}
\caption{ The effect of EoS on the disk mass.  $M_0$ is the total baryonic mass outside the BH. 
}
\label{fig:eos_mass_effects}
\end{center}
\end{figure}

\begin{figure}
\begin{center}
\includegraphics[width=9cm]{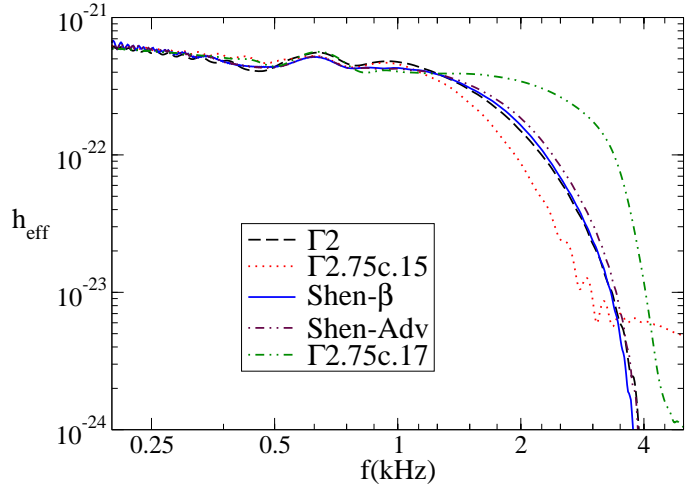}
\caption{
  The effective amplitude, as defined in Eq.~(41) of~\cite{Shibata:2009cn},
  for each run.  The
  assumed distance from the source is 100Mpc. 
}
\label{fig:eos_spectrum_effects}
\end{center}
\end{figure}

The effects of the EoS on the gravitational wave spectrum and on the
post-merger disk mass are shown in figures~\ref{fig:eos_mass_effects}
and~\ref{fig:eos_spectrum_effects}.  The leading contribution from the
EoS is related to the compaction of the star.  As Shibata {\it et
  al.}~\cite{Shibata:2007zm,Shibata:2009cn} found for $\Gamma=2$
stars, we observe that a higher compaction leads to stronger
gravitational waves and a smaller disk. The more
compact star reaches the ISCO with minimal distortion and falls nearly
entirely into the black hole, with only a small fraction of its mass
sent out in a relatively short tidal tail extending to $d\sim$250km.
The tail forms a small disk of mass $\approx 0.02M_{NS}$.  The cutoff
frequency of the gravitational waves ($f\sim2$kHz) is higher than for
more extended stars ($f\sim1$kHz for $\Gamma$2.75c.15).

By comparison, modifying the EoS while keeping the compaction constant seems to have more modest
effects.
For NS compaction of $\mathcal{C}=0.15$, every EoS predicts a final disk mass $\sim 0.05-0.1M_{\rm NS}$, 
i.e. $M_{\rm disk}\approx 0.1M_{\odot}$.  The density-weighted average temperature of the
disk is 1-2MeV, although some regions in the disks reach $T\approx 20$MeV.
Differences in the gravitational wave signal are more visible: the $\Gamma=2.75$ star has a lower cutoff
frequency and peak amplitude. But even then, the differences remain smaller than those due to the
compactness of the star.  The close similarity of the Shen and $\Gamma=2$ waveforms is surprising. 
A partial explanation may lie in the fact that the Shen-EoS is
actually very soft at the densities of the matter after the NS is disrupted.  In these regimes, Shen
is closer to $\Gamma=2$ than to $\Gamma=2.75$.

Another strong effect of the stiffness of the EoS is the different behaviors of the tidal tails, which can lead to important effects on the post-merger disk dynamics.
For all EoS, our simulations end with $0.06-0.07M_{\rm NS}$ of matter within 200km of the BH
 --- either in the disk or in the process of joining it.  The mass and
 size of the tidal tail, however, vary: for the $\Gamma=2.75$ star,
 the tail mass is $0.06M_{\rm NS}$ and more than 2\% of the matter
 will eventually go as far as 2000km away from the BH before falling
 back on a timescale of 200ms (assuming geodesic motion).  When the tail
is maximally extended, our grid would be unable to adequately resolve the
whole system.  Therefore, we stop evolving when the tail is still expanding. 
On the other
 hand, the $\Gamma=2$ EoS leads to a much smaller tail, with a mass of
 $0.02M_{\rm NS}$. Only a negligible amount of matter reaches a
 distance of 500km, and most of the matter would fall back on the disk
 within 25ms of the disruption. For both polytropes, interactions
 between the disk and the tail are strong enough to keep the disk from
 settling to an axisymmetric state over the duration of our
 simulation.  The Shen case also has a tail of mass
 $0.02M_{\rm NS}$, but it is thicker and falls back more rapidly
 onto the disk.  The disk will settle more quickly, and is much closer
 to axisymmetry at the end of our simulation.

For both Shen-$\beta$ and Shen-Adv runs, the density-weighted average of $Y_e$
remains close to $\langle Y_e\rangle=0.09$ throughout the inspiral, so the
inspiral is essentially identical for the two runs.  The mergers are also
similar, although the Shen-$\beta$ waveform decays somewhat more rapidly. 
The composition of the disk,
however, is radically different.  For Shen-Adv, the final
$\langle Y_e\rangle$ is about 0.09.  (It decreases slightly, even though
there are no $Y_e$ source terms, because the more highly-leptonized
central region of the NS is swallowed by the BH.)  The baryon mass is
about 86\% free neutrons, 7\% heavy nuclei, 5\% free protons, and 2\%
alpha particles.  The nuclei have an
average $\langle A\rangle\approx 95$, $\langle Z\rangle\approx 30$.  
If $\beta$-equilibrium is enforced, $\langle Y_e\rangle$ increases to 0.2 as
the NS matter decompresses.  The baryon mass is about 65\% free neutrons,
25\% heavy nuclei ($\langle A\rangle\approx 75$, $\langle Z\rangle\approx 30$),
9\% free protons, and 1\% alpha particles.  The disks produced by the
Shen-$\beta$ and Shen-Adv mergers have similar densities and temperatures: 
the density-weighted average temperature is about 2.5MeV for Shen-Adv and
2.7MeV for Shen-$\beta$, although in each case the maximum temperature reaches
$\approx 12$MeV, and the average is slowly increasing.

\subsection{The final black hole and disk state}
\label{final}

The final state of the black hole is given in
Table~\ref{table:final_BH}.  The final spin $s_{\rm final}$ is in the
range 0.7-0.8, with the Shen runs having higher $s_{\rm final}$
because they merge somewhat more quickly and have less time to radiate
angular momentum.  The BH kick velocity is $\sim$10-100km s${}^{-1}$

\begin{figure}
\begin{center}
\includegraphics[width=9cm]{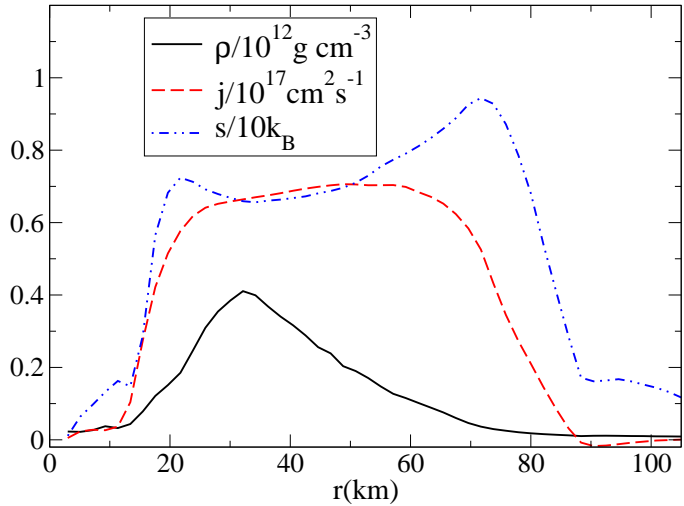}
\caption{ 
  The density $\rho$, specific angular momentum $j$, and
  specific entropy $s$ of the Shen-$\beta$ disk as a
  function of cylindrical radius $r$, shown 10ms ($370M$) after
  the merger. Each point is a $\rho$-weighted
  average over the angular and vertical directions.
}
\label{fig:profiles}
\end{center}
\end{figure}

In figure~\ref{fig:profiles}, we plot the profiles of the density
$\rho$, specific entropy $s$, and specific angular momentum
$j=u_{\phi}/u_{\rm t}$ for run Shen-$\beta$ at final time $t=$10ms
after the merger.  By 3ms after merger begins, a distinct torus
forms around the BH.  By 5ms after merger, matter begins to clear
out in the region near the hole, and the accretion rate $\dot{M}_{\rm disk}$ drops to a low
value ($\tau=M_{\rm disk}/\dot{M}_{\rm disk}\approx 24$ms).  The torus has a maximum density
of $\rho=10^{12}$g cm${}^{-3}$
located at a coordinate radius of $r=r_c=30$km from the BH.  The width
of the torus is $\approx20$km, and the height is $\approx10$km.  At
this time, the disk itself has not yet stabilized.  The gas more than
20km from the BH has nearly constant $s$ and $j$, while the gas nearer
the hole has somewhat higher $s$ and much lower $j$.  Subsequent motions in
the fluid lead to modest but positive radial gradients for $s$ and $j$
in most of the high-density region, the exception being the
small but negative entropy gradient that persists around $r\approx 30$km. 
In the still-settling, lower-density
outer regions, $s$ and $j$ drop significantly.  The angular velocity
$\Omega$ decreases with $r$ everywhere: $\Omega\sim r^{-1.2}$ in the
high-density region.
The assumption of $\beta$-equilibrium is probably not good in the
outer regions of the disk, which are optically thin to neutrino
emission.  The profiles of $\rho$ and $j$ are quite similar for the
Shen-Adv disk, so these variables appear to depend weakly on $Y_e$.

The longer-term evolution of the disk depends on physical processes
not included in these simulations.  Since $d\Omega/dr<0$, our disks
are subject to the magnetorotational instability (MRI)~\cite{1998RvMP...70....1B}. 
Turbulence induced by this instability might have the effect of an $\alpha$-viscosity of magnitude
$\alpha\sim 0.01-0.1$~\cite{2007PThPh.118..257S}.  This will heat the disk,
redistribute angular momentum, and drive accretion.  The disk's only significant
cooling mechanism is neutrino emission~\cite{1999ApJ...518..356P,
2007ApJ...657..383C}. 
3D Newtonian simulations predict that a neutrino-cooled disk with $\alpha\sim 0.1$ and
$M_{\rm disk}$ similar to ours will accrete on
a timescale of $\sim$0.1 second and release energy in neutrinos at a rate
$L_{\nu}\sim 10^{53}$erg s${}^{-1}$, possibly making such a merger remnant a viable GRB
candidate~\cite{2006A&A...458..553S}.

\section{Conclusions}
\label{conclusions}
We have investigated the influence of the equation of state on
BHNS binaries.  We find that the NS compaction
has a strong influence on the disk mass and the cutoff frequency
of the gravitational waveform.  The effects of EoS stiffness for a
fixed compaction are weaker.  The overall behavior of
the merger was found to be independent of EoS stiffness---we
never find episodic mass transfer or unbound ejecta.  The stiffness of
the EoS in the lower-density outer layers of the NS does affect the
merger dynamics, with stiffer EoS leading to larger and longer-lived tidal
tails, and this effect manifests itself in the merger waveform.  We
find that the evolution of $Y_e$ weakly influences the waveform and the
final disk mass. 

Our simulations suffer from two classes of limitations.  First, we have
considered only a very small sample of the interesting BHNS parameter space.  It
is possible that the effects of EoS stiffness are more pronounced at
different binary mass ratios or different BH spins.  Also, we have
considered only three EoS. 
A more systematic approach would be to use a single EoS with
adjustable parameters (e.g.~\cite{Read:2009yp,Lattimer:1991nc}). 
An adequate EoS for this purpose would have to cover the range of
likely NS stiffness and radius while capturing all of the important
physical EoS features, including (for the post-merger evolution)
its complicated temperature and composition dependence. 
Other important areas for
improvement involve our treatment of the NS microphysics.  To evolve
the final disks realistically, the effects of neutrino radiation on
the temperature and composition of the matter must be included.  The
evolution of the disk is also strongly affected by the presence of
magnetic fields, and in particular by the MRI.  Simulations that
include radiation and magnetohydrodynamics are needed to assess the
ability of these disks to produce GRBs.

\ack

We thank Evan O'Connor, Harald Pfeiffer, and Manuel Tiglio for useful discussions. 
This work was supported in part by a grant from the Sherman Fairchild
Foundation, by NSF grants PHY-0652952 and PHY-0652929, and NASA grant
NNX09AF96G.  CDO is partially supported through NSF award
No. AST-0855535.  This research was supported in part by the NSF through
TeraGrid~\cite{teragrid} resources provided by LONI's Queen Bee and
NCSA's Ranger clusters.  Computations were also performed on the GPC
supercomputer at the SciNet HPC Consortium.  SciNet is funded by: the
Canada Foundation for Innovation under the auspices of Compute Canada;
the Government of Ontario; Ontario Research Fund - Research Excellence;
and the University of Toronto.  We thank Harald Pfeiffer for getting us
access to SciNet by compiling SpEC and submitting our runs there.

\section*{References}
\bibliographystyle{iopart-num}
\bibliography{References/References}

\end{document}